\documentclass[12pt]{iopart}

  \usepackage{epsfig,graphics,graphicx}
  \usepackage{slashed}
  \usepackage{cite}
  \usepackage[latin1]{inputenc}
  \usepackage{mathrsfs}
\usepackage{mathcomp}
\usepackage{yfonts}

\usepackage{bm,mathrsfs,slashed}

\long\def\comment#1{ }

\newcommand{\eqn}[1]{Eq.~\ref{#1}}

\newcommand{\beq}{\begin{eqnarray}}
\newcommand{\eeq}{\end{eqnarray}}

\newcommand{\bq}{{\bar q}}

\newcommand{\tqq}{\theta_{q\bar q}}
\newcommand{\tq}{\theta_{q}}
\newcommand{\tbq}{\theta_{\bar q}}
\newcommand{\tf}{\theta_{f}}
\newcommand{\ts}{\theta_{s}}

\def\tlambda{\tau_{\lambda}}

\begin{document}

\title[Interference Phenomena  in Medium Induced Radiation]{Interference Phenomena in Medium Induced Radiation}

\author{J.~Casalderrey--Solana,$^{\,a}$  E. Iancu,$^{\,a,b}$}
        
\address{${}^a$CERN, Theory Division, CH-1211 Geneva, Switzerland
\\
 ${}^b$Institut de Physique Th\'eorique,
CEA Saclay,
 F-91191 Gif-sur-Yvette, France
}
\ead{jorge.casalderrey@cern.ch, edmond.iancu@cea.fr}
\begin{abstract}
We consider the interference
pattern for the medium--induced gluon radiation produced by a color
singlet quark--antiquark antenna embedded in a QCD medium with size $L$ and 
`jet quenching' parameter $\hat q$. Within the 
BDMPS--Z regime, 
we demonstrate that, for a dipole opening angle  $\theta_{q\bar q} \gg\theta_c\equiv {2}/{\sqrt{\hat q L^3}}$, the interference between the
medium--induced gluon emissions by the quark and the antiquark is
suppressed with respect to the
direct emissions. This is so since 
direct emissions are delocalized  throughout the medium
and thus yield contributions proportional to $L$ while 
interference occurs only between emissions
at  early times, 
when both sources remain coherent.
Thus,
for $\tqq \gg\theta_c$, the medium--induced radiation
 is the sum of the two spectra
individually produced by the quark and
the antiquark, without coherence effects like angular ordering.
For $\tqq \ll\theta_c$, the medium--induced radiation 
vanishes.

\end{abstract}

%\maketitle
\section{Introduction}
One of the most spectacular observations of the LHC heavy ion program  is the strong 
modification of jets in a dense QCD medium \cite{cmsjets,atlasjets}. Most  theoretical descriptions
of this modification have focussed on the characterization of the medium--induced gluon radiation off a single parton propagating through the plasma (see \cite{CasalderreySolana:2007zz} for a review). However,  due to both jet evolution and in--medium radiation, a jet involves several partons which can radiate simultaneously. It is then necessary to address whether
the medium--induced emissions by several sources in such a multi--partonic system can interfere with each other.
In vacuum, interference effects are  present, frustrating large angle emissions and leading to the {\em angular ordering} of vacuum showers. The survival of angular ordering for medium--induced radiation has only recently been discussed \cite{CasalderreySolana:2011rz, MehtarTani:2010ma}.

To address this problem, we will describe  the medium--induced gluon radiation
from a  $q\bq$ dipole created in the medium. If the dipole angle is sufficiently large, 
 the radiation produced by the $q$ and the
$\bar q$ has no overlap with each other
and thus they are independent. 
If the dipole opening angle is not that large, and in view of the experience with radiation in the vacuum, one may
expect the dipole antenna pattern to be affected by interference effects.
However,
 in the medium this expectation is generally incorrect. 
 
A detailed proof of the of the statement above can be found in \cite{CasalderreySolana:2011rz} and
relies on the following properties
of the medium--induced radiation by a single source: 
(i) The formation of a medium--induced gluon takes a time
    $\tau_f=\sqrt{2 \omega/\hat q}$. When the gluon is formed it is emitted at a typical
    angle $\tf=\left(2 \hat q/\omega^3\right)^{1/4}$ from the parent quark and it carries a typical
    transverse momentum $k_f\simeq\omega \tf$.
(ii) After formation,  multiple scattering leads to  additional
    broadening of the gluon spectrum. The final gluon distribution is
    concentrated within a typical  angle $\ts=\sqrt{\hat q L}/{\omega} >\tf$ around the parent
    parton.
(iii) As long as $\tau_f\ll L$, gluons are emitted all along the
    medium  and the medium--induced gluon
    spectrum is proportional to $\tau_f L$ (the longitudinal phase--space).

\section{Coherence phenomena and interference  time scales}
The interference phenomena require the two
partonic sources 
 to be {\em coherent} with each other
during the gluon formation. In turn, this requirement has {\em two}
aspects:

\texttt{(i) Quantum coherence.} The emission process preserves the
quantum coherence of the $q\bq$ system so long as the virtual gluon
overlaps with both sources  during formation.
 For that to be
possible, the transverse resolution of the gluon at the time of
formation, as measured by the respective transverse wavelength
$\lambda_f= 1/k_f$, should be larger than the typical distance between
the $q$ and the $\bar q$. In the
vacuum, this condition  leads to angular ordering but for medium--induced emissions
the situation is more subtle: during formation, the transverse
size of the $q\bq$ system increases from $r_{min}\simeq\tqq\, t_1$ to
$r_{max}\simeq\tqq\, t_2$~;
here, $t_1$ and $t_2=
t_1+\tau_f$ are the times when the emission is initiated and completed respectively. 
Since the gluon undergoes transverse diffusion during
its formation, the relevant  size is the 
 {\em geometric average} of these two scales and 
 the condition of quantum coherence amounts to
$\sqrt{r_{min}r_{max}} < \lambda_f$, or
$\sqrt{t_1(t_1+\tau_f)}\ < \lambda_f/\tqq\,\equiv\,
 \tau_\lambda$
with $\tlambda$ the 
{\em transverse resolution time}.
  It is easy to see that
 \beq \label{tlambda}
 \tau_\lambda\,=\,\frac{1}{\tqq\,(\hat q\omega)^{1/4}}
 \,=\,
\tau_f\, \frac{\tf}{\tqq}\,,
 \eeq
where the second estimate follows from $k_f\simeq \omega \theta_f$ and
$\tau_f\simeq 2/(\omega\tf^2)$.
This estimate leads to  
two limiting regimes,
depending upon the ratio $\tqq/\tf$ :

\texttt{(i.a)} For relatively small dipole angles $\tqq\ll\tf$,  one has
$\tlambda\gg\tau_f$ and 
the definition of $\tau_\lambda$ 
implies $t_1<
\tlambda$. 
Since in this regime the BDMPS--Z spectra produced by the two
emitters are confined to angles $<\tf$ around their respective
sources during the formation process, these two
spectra overlap  but can only interfere over a time $\tau_\lambda$.

\texttt{(i.b)} For larger dipole angles $\tqq\gg\tf$, one has
$\tlambda\ll\tau_f$ and therefore $t_1\ll\tau_f$ as well.
The definition of $\tau_\lambda$ leads to
$t_1\,< \tlambda^2/\tau_f\,\equiv\,\tau_{int}$
 %\beq\label{tintdef}
% t_1\,< \frac{\tlambda^2}{\tau_f}\,\equiv\,\tau_{int}
%  \,,\eeq
which introduces a new scale $\tau_{int}$, the {\it interference time}.
This scale can be rewritten as
 \beq \label{tint1} \tau_{int}=\frac{2}{\omega \tqq^2}=\tau_f
\left(\frac{\tf}{\tqq}\right)^2\,,
 \eeq
where the first expression is recognized as the {\em vacuum--like}
formation time for a gluon emitted at an angle $\sim\tqq$. 
This is so since, in this regime,
 interference can only occur for  gluons radiated at this large angle. 

\texttt{(ii) Color coherence.} In addition to quantum coherence, the
existence of interference effects 
require the preservation of the color coherence between the quark and the
antiquark. In the vacuum, the color state of the dipole is conserved
until a gluon emission takes place and the interference pattern  is
governed solely by quantum coherence. In the medium, on the contrary, the
interactions with the medium constituents change the color of each of the
propagating partons via `color rotation'. 

For a very energetic parton,
this rotation amounts to multiplying the respective wavefunction by a
a Wilson line.
For the $q\bq$ pair we have two such Wilson lines which diverge from each
other  at constant angle $\tqq$.
The color coherence is measured by the 2--point correlation function of
these Wilson lines, as obtained after averaging over the fluctuations of
the background field.  Within the `multiple soft scattering
approximation', this 2--point function can be computed 
and it 
shows that the quark and the antiquark
loose any trace of their original color state after the {\it
decoherence time}
 \beq\label{tdecoh}
 \tau_{coh}\,=\,\frac{2}{
 (\hat q\tqq^2)^{1/3}}\,
 =\tau_f \, \left(\frac{\tf}{\tqq}\right)^{2/3}
\,=\,\left(\frac{\theta_c}{\tqq}\right)^{2/3} L .\eeq

\section{Classification of dipole sizes.}

\begin{figure}[tb]
\begin{center}
\includegraphics[width=0.5\textwidth]{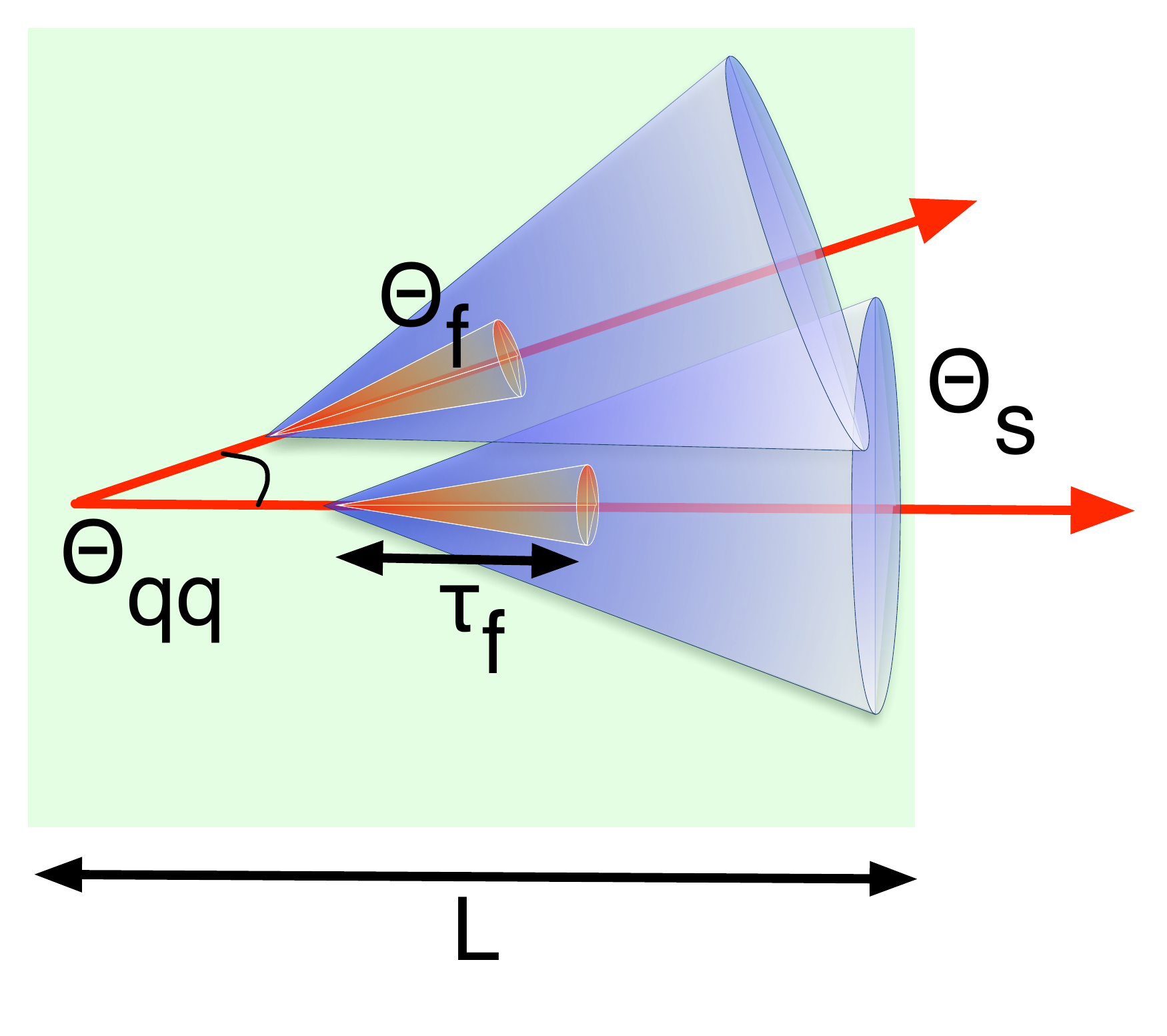}
\includegraphics[width=0.40\textwidth]{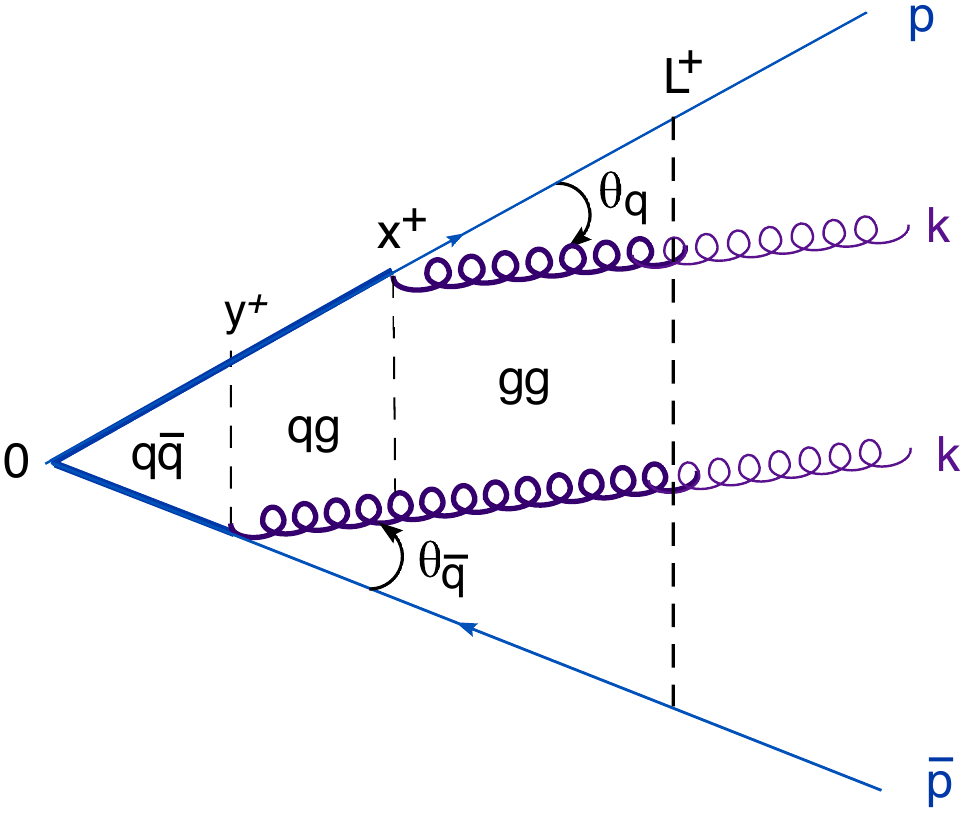}
\end{center}
\caption{\label{fig:medium}\sl Left: Sketch of the radiation fronts for a relatively large dipole (see text for definition). While the final radiation spectra from each source overlap (blue cone), they are separated at formation (red cone).
 %Interference is only possible a very early times. 
 Right: Feynman graph for interference. The amplitude and complex conjugate amplitude are drawn on top of each other. }
\end{figure}

While the details of the in--medium dipole antenna pattern 
 depend upon all the scales identified above, the phase--space for interference is
controlled by the {\em smallest} of them,
$\tau_{min}=\min(\tau_\lambda,\tau_{int},\tau_{coh})$.
As a consequence, the
interference contribution to the gluon spectrum,
 does not scale with the medium length, as the emission
from each of the sources does, but with $\tau_{min}$. 
%By inspection of the interference time scales
 We can distinguish the following regimes  \cite{CasalderreySolana:2011rz}

\texttt{1. Very large dipole angles, $\tqq >
\ts$.}  In this case the medium-induced spectra from each source do not overlap
and the $q$ and the $\bar q$ radiate independently.

\texttt{2. Relatively large dipole angles, $\tf< \tqq <
\ts$.} In this regime Eqs.~\ref{tlambda}, \ref{tint1} and
\ref{tdecoh} imply the hierarchy of scales
 %\beq
 $   \tau_{int}< \tau_\lambda <  \tau_{coh}
 <  \tau_f.
 $
Accordingly, in this regime, the longitudinal phase--space for
interferences is of order $\sim\tau_{int}\tau_f$ and it is suppressed
with respect to the corresponding phase--space $\sim\tau_f L$ for direct
emissions by a factor
 \beq\label{Rint}  \mathcal{R}=\frac{\tau_{int}}{L}
  \sim \sqrt{\frac{\omega}{\omega_c}}
 \left(\frac{\tf}{\tqq}\right)^2  \ll 1\,.
 \eeq

%\vspace*{0.2cm} \noindent

\texttt{3. Relatively small dipole angles $\theta_c\ll \tqq\ll\tf$.} In
this case, the  limitation on the phase--space for interference
is due to color coherence, since the  ordering of time scales is reverted:
$
\tau_f \ll \tau_{coh} \ll \tau_\lambda \ll \tau_{int}
$.
The longitudinal phase--space for interference is now of order
$\tau_{coh}\tau_f$ and it is 
suppressed as compared to the phase--space for direct emissions. Hence the interference
contribution is suppressed by 
 \beq  {\mathcal R}\,=\,\frac{\tau_{coh}}{L}\,=
 \left(\frac{\theta_c}{\tqq}\right)^{2/3}\ll\,1\,.
 \eeq
Note that, in this case, the medium--induced radiation 
is distributed at large angles $\tq\simeq
\tbq>\tf\gg\tqq \,, $ well outside the dipole cone, and  one may
wonder why the total radiation is not zero.
The reason is that, so long as $\tqq\gg\theta_c$, a $q\bq$ pair
immersed in the medium is {\em not} a `color singlet' anymore, except for
a very brief period of time $\sim \tau_{coh}$.

\texttt{4.  Very small dipoles angles $\tqq< \theta_c$.} 
For these small angles, the color
coherence time $\tau_{coh}$ becomes as large as the medium size $L$, as
clear from 
\eqn{tdecoh}, and the $q\bar q$ pair preserves its
color and quantum coherence throughout the medium. Interference
effects are not suppressed  and they act towards reducing the
medium--induced radiation by the dipole. For sufficiently small angles
$\tqq\ll\theta_c$, the color decoherence is parametrically small and the
total in--medium radiation becomes negligible.

{\bf In summary}, we have argued that for  dipole angles
$\tqq\gg\theta_c$, the interference effects for the medium--induced
radiation are negligible, so the total BDMPS--Z spectrum by the dipole is
the incoherent sum of the  spectra produced by the $q$ and
the $\bar q$. For smaller angles $\tqq<\theta_c$, the
interference effects are not suppressed and they 
cancel the direct emissions when $\tqq\ll\theta_c$.  We observe that for
the representative values for $\hat q=10\, {\rm GeV^2/fm}$ and $L=10\, {\rm fm}$,
$\theta_c\sim 0.01$ is very small, and most of the dipoles of phenomenological interest will radiate as two independent partonic sources. This simplifies  the way towards Monte--Carlo studies of the in--medium jet evolution.
%The transition
%between the two regimes, occurring at $\tqq\simeq\theta_c$, could in
%principle be studied within the formalism that we shall develop later.
%However, such a study goes beyond the approximation schemes that we shall
%use throughout this paper and which are adapted to the most interesting
%regime at $\tqq\gg\theta_c$.

\end{document}